\documentclass[12pt]{iopart}

\usepackage{iopams}

\usepackage{amsfonts}
\usepackage{amsmath}
\usepackage{graphicx}
\usepackage{bm}
\usepackage{fancyhdr}
\usepackage{float}
\usepackage{cases}
\usepackage{hyperref}
\usepackage{graphicx,epstopdf}   
\usepackage{booktabs} 
\usepackage{xcolor}

\begin{document}

\title{Landau free energy and the absence of spontaneous magnetization of the one-dimensional Ising model}

\author{Zheng Z F$^{1, 2}$, Lin R K$^{1,2,3}$ and Zhang J M$^{1,2,*}$}

\address{$^1$ Fujian Provincial Key Laboratory of Quantum Manipulation and New Energy Materials, College of Physics and Energy, Fujian Normal University, Fuzhou, China}
\address{$^2$ Fujian Provincial Collaborative Innovation Center for Advanced High-Field Superconducting Materials and Engineering, Fuzhou, China}
\address{$^3$ State Key Laboratory of Functional Crystals and Devices, Fujian Institute of Research on the Structure of Matter, Chinese Academy of Sciences, Fuzhou, China}

\vspace{10pt}
\begin{indented}
\item[]$^*$ Author to whom any correspondence should be addressed.
\end{indented}

\ead{wdlang06@gmail.com}

\begin{abstract}
We revisit the problem of spontaneous magnetization of the one-dimensional Ising model from the Landau free energy perspective. To this end, we define and calculate the density of states of the one-dimensional Ising model following a technique introduced by Ising. The observed monotonicity property of the density of states  suggests heuristically that the model does not exhibit spontaneous magnetization at any finite temperature. Subsequently, we solve the model exactly in the thermodynamic limit by employing the maximum-term approximation, which is feasible due to the simple analytical expression of the density of states.
We also show that the Landau free energy is an increasing function of $|m|$ and its second derivative at $m=0$ is positive and non-analytic in temperature, proving rigorously the absence of spontaneous magnetization of the model at any finite temperature. 
\end{abstract}

\submitto{\EJP}
\maketitle

\section{Introduction}

It is well-known that the one-dimensional (1D) Ising model does not show spontaneous magnetization at any finite temperature. Actually, this conclusion was first reached by Ising himself, in his very original paper studying the model named after him \cite{Ising1925, english}. Ising arrived at this conclusion by solving the model exactly. 

The lack of spontaneous magnetization of the 1D Ising model is nowadays generally understood via a heuristic argument credited to Landau \cite{peliti, landaulifshitz}. In the ground states, all the $L $ spins are parallel, either all upwards or all downwards. However, the system possesses $\mathcal{O}(L)$ configurations containing a single domain wall. These domain wall configurations suffer an energy penalty of $\mathcal{O}(1)$ but significantly alter the system's magnetization. Therefore, at any finite temperature, as long as the lattice size $L $ is large enough, configurations with a domain wall are overwhelmingly more probable than the fully polarized ground states. In the literature, this is often succinctly described as ``entropy wins over energy''.

In this paper, we revisit the problem from a perspective which also goes back to Landau \cite{landaulifshitz, landau,  russian}. In a lecture given by the last author, an undergraduate student, who was unsatisfied with the explicit but unintuitive Ising solution and the insightful but  ``vague'' argument above, posed the question: At a given temperature $T $, what value of the magnetization $M$ is the most probable? Admittedly, this is a very natural question concerning the problem of spontaneous magnetization, since it goes directly to the probability distribution of $M$ and as is well known, in the thermodynamic limit, the most probable value is often also the mean value. As each configuration has a definite value of energy (and thus a corresponding Boltzmann weight) and a definite value of magnetization, the probability of the magnetization being $M$ is proportional to the partial partition function $Z( T,M)$. Here, $Z(T,M)$ is the sum of the statistical weights of all configurations restricted to magnetization $M$, and summing it over all $M$ gives the full partition function $Z(T)$. The concern is then whether $M = 0$ is a local maximum or a local minimum of $Z( T,M)$, with the latter corresponding to spontaneous magnetization. In terms of free energy, we note that ($k_B = 1 $ throughout this paper) $-T \ln Z(T,M)= F( T,M)$ is the Landau free energy, and the problem reduces to determining whether $M=0$ is a local minimum or a local maximum of $F$.   

It is interesting that the question above can be answered almost without calculation. Strictly speaking, this requires some priori knowledge of the density of states of the model. 
Following Ising, we can obtain explicit expressions of the density of states of the model. Unexpectedly, we found that it is essentially monotonic. An immediate consequence is that, at any finite temperature, the Landau free energy $F( T, M )$ is a monotonically increasing function of $|M|$, which rules out the possibility of spontaneous magnetization. 

We emphasize that the absence of spontaneous magnetization in the 1D Ising model is a well-established result. The purpose of this paper is not to present a new physical discovery, but rather to offer a novel pedagogical reformulation. By bridging Ising's original combinatorial counting with the framework of Landau free energy, we aim to provide an instructive, alternative perspective for students of statistical mechanics.

\section{The model}

We consider a 1D Ising model with the open boundary condition. The system consists of a chain of $L $ spins $s_i $, $1\leq i \leq L $. Each spin takes either value $+1$ or $-1 $. There are thus in total $2^L $ configurations. The energy of a configuration $\vec{s} = (s_1, s_2, \ldots, s_L )$ in an external magnetic field of strength $h $ is
\begin{eqnarray}
  \mathcal{E}_{total} &=& E - h M , 
\end{eqnarray}
where $E$ is the interaction energy,
\begin{eqnarray}\label{eint}
  E &=& - \sum_{i=1}^{L-1 } s_i s_{i+1 } ,
\end{eqnarray}
and $M $ is the magnetization of the system, 
\begin{eqnarray}
  M &=&  \sum_{i=1}^{L } s_i .  
\end{eqnarray}

We classify the configurations according to the values of $(E, M )$, and introduce the notion of density of states (or number of states) $\mathcal{D}(E, M)$, which is the number of configurations with interaction energy $E$ and magnetization $M$. 
Apparently, $E$ and $M $ take a finite number of integral values. Specifically, $E\in \{-L+1, -L+3, \ldots, L-1\}$, and $M \in \{-L, -L+2, \ldots , L \}$. See Table \ref{tabledos} for an example. 

At a given temperature $T$, the partition function $Z( T, h )$ can be expressed in terms of the density of states ($k_B = 1 $, $\beta \equiv  1/T $, $\alpha \equiv  \beta h $),
\begin{eqnarray}\label{z}
  Z(T, h ) &=& \sum_{E, M} \mathcal{D}(E, M) e^{-\beta E +\alpha M  } \equiv \sum_{E, M} W(E, M ).  
\end{eqnarray}
We also introduce the partial partition function 
\begin{eqnarray}\label{partialz}
  Z(T, h, M ) &=& \sum_{E} \mathcal{D}(E, M) e^{-\beta E +\alpha M  } \equiv  \sum_E W(E, M ),
\end{eqnarray}
where we only consider the configurations with magnetization $M $. The significance of the partial partition function $Z(T, h, M )$ is that, the ratio $Z(T,h, M)/Z(T, h )$ is the probability of system having magnetization $M $. Like the logarithm of the partition function is the free energy, the logarithm of the partial partition function is the Landau free energy, 
\begin{eqnarray}
  F(T,h, M) &=& -T \ln Z(T, h, M) . 
\end{eqnarray}
We emphasize that, unlike the phenomenological Landau theory typically presented in textbooks—which relies on an approximate polynomial expansion—we shall calculate the Landau free energy as a constrained partial free energy exactly from the microscopic Hamiltonian above.

\begin{figure*}[tb]
\includegraphics[width= 0.8\textwidth ]{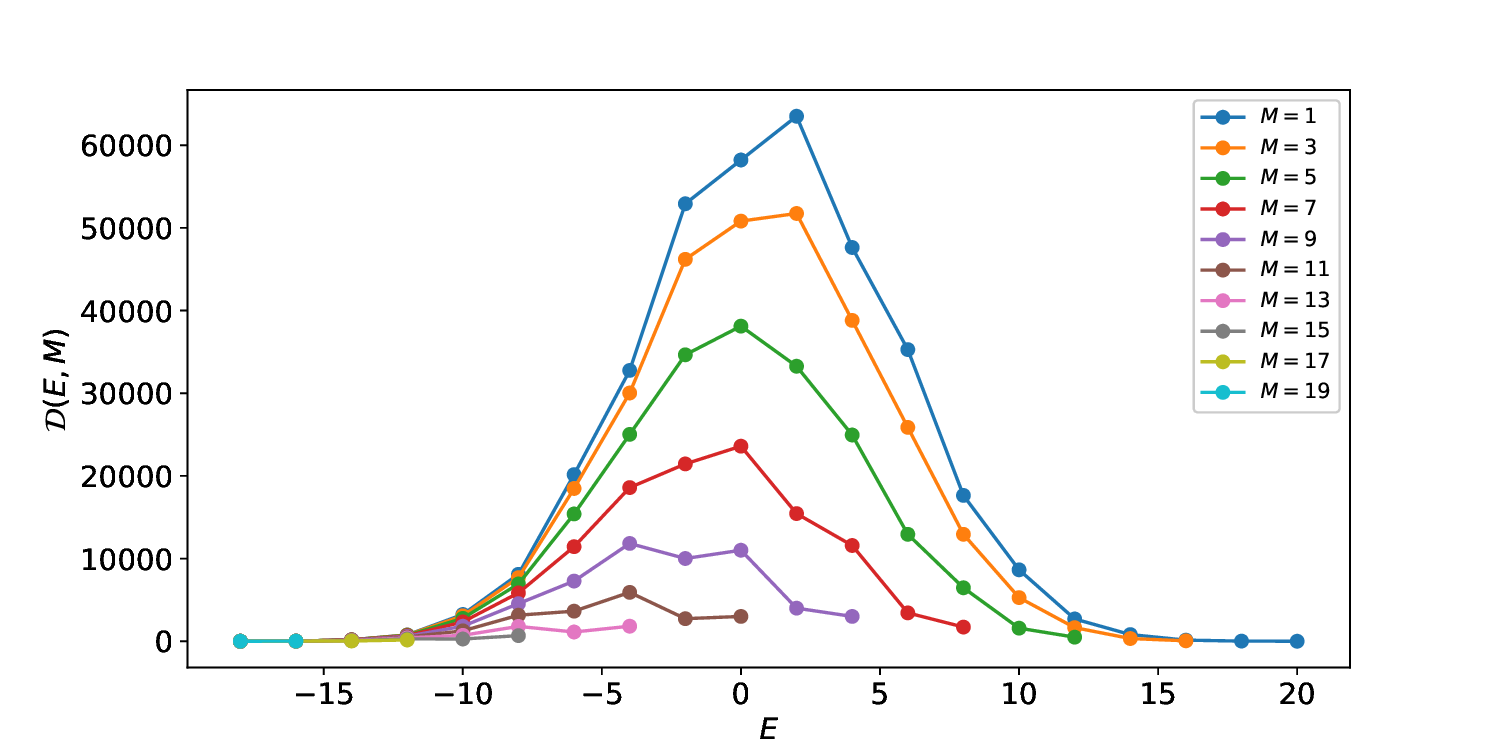}
\caption{(Color online) Density of states $\mathcal{D}(E, M) $ of a 1D Ising model with the open boundary condition. Only nonzero values are shown.  The total number of spins is $L = 21$. Each curve is for a different value of the magnetization $M $.  We see that the curve corresponding to $M+2$ is always below that to $M $, in accord with (\ref{indos}).  }
\label{fig1}
\end{figure*}

\begin{table}[h]
\centering
    \resizebox{\textwidth}{!}{%
        \begin{tabular}{c|ccccccccccccccc}
            \toprule
            $E \setminus M$ & \textbf{-14} & \textbf{-12} & \textbf{-10} & \textbf{-8} & \textbf{-6} & \textbf{-4} & \textbf{-2} & \textbf{0} & \textbf{2} & \textbf{4} & \textbf{6} & \textbf{8} & \textbf{10} & \textbf{12} & \textbf{14} \\
            \midrule
            \textbf{-13}  & 1 &  &  &  &  &  & &  &  &  &  &  &  &  & 1 \\
            \textbf{-11}  &  & 2 & 2 & 2 & 2 & 2 & 2 & 2 & 2 & 2 & 2 & 2 & 2 & 2 &  \\
            \textbf{-9}  & & 12 & 12 & 12 & 12 & 12 & 12& 12& 12 &12 &12 & 12 & 12 & 12 &  \\
            \textbf{-7}  &  &  & 22 & 40 & 54	&64	&70&72	&70	&64	&54	&40	&22	 &  &  \\
            \textbf{-5}  &  &  & 55&	100	&135&	160&	175&	180&	175&	160	&135	&100	&55 &  &  \\
            \textbf{-3}  &  &  &  & 90&	216&	336	&420	&450	&420	&336	&216	&90 & &  & \\
            \textbf{-1}  &  &  &  & 120	&288&	448&	560&	600&	560	&448	&288	&120 & &  &  \\
            \textbf{1}  & & &  & & 168	&448&	700&	800&	700&	448	&168& & &  & \\
            \textbf{3}  & &  &  &  & 126	&336	&525&	600&	525&	336	&126 & &  &  &  \\
            \textbf{5} &  &  &  &  &  & 140	&350&	450&	350&	140 & & &  &  &  \\
            \textbf{7} &  &  &  & & &56&	140	&180	&140	&56 &  & &  & &  \\
            \textbf{9} & &  &  &  & &  & 42	&72&	42 &  &  & & & &  \\
            \textbf{11} &  &  &  & &  & &7	&12&	7 & & & &  & & \\
            \textbf{13} &  &  &  &  &  &  &  & 2 &  & &  &  &  &  &  \\
            \bottomrule
        \end{tabular}%
    }
\caption{Density of states $\mathcal{D}(E, M) $ of a 1D Ising model with the open boundary condition. Only nonzero values are shown.  The total number of spins is $L = 14$. One can check that except for the first row, in all other rows, $\mathcal{D}$ is even in $M$, peaks at $M=0$, and decreases monotonically with $|M|$, in accord with (\ref{indos}). }
\label{tabledos}
\end{table}

\section{Density of states and absence of spontaneous magnetization }

We need knowledge of the density of states $ \mathcal{D}(E, M)$ to calculate or compare the (partial or full) partition functions. First, it is manifestly symmetric in $M$, i.e., $\mathcal{D}(E, M) = \mathcal{D}(E, -M)$. Second, for the extremal values of $M$, $M=\pm L $, the spins are all aligned, $\mathcal{D}$ is nonzero only when  $E= -(L-1)$. For other values of $(E, M)$, we follow Ising. For given $(E,M) $, the numbers of upper spins and down spins are $N_{\uparrow} = (L+ M)/2$ and $N_{\downarrow} = (L - M)/2$, respectively,  and the number of domain walls is $N_{dw} = (E+L-1 )/2$, since each domain wall contributes an energy penalty of 2 and the energy of the ground state (which is free of domain walls) is $-(L-1)$. The number of domains is then $N_{dw}    +1 $.  As long as $M\notin \{ -L, L \}$, $N_{dw}  \geq 1 $. 

If $N_{dw}  $ is odd, i.e., $N_{dw}  = 2 K - 1 $ for some positive integer $K $, there are $K $ upper domains and $K $ down domains. The total number of configurations is then 
\begin{eqnarray}\label{dosodd}
  \mathcal{D}(E, M) &=& 2  \binom{N_\uparrow - 1 }{K -1 } \binom{N_\downarrow -1 }{K  -1 },
\end{eqnarray}
The factor 2 comes from the fact that the lattice can either start with an upper domain or a down domain. The binomial factors count the ways of decomposing the number $N_\uparrow$ or the number $N_\downarrow$ into $K $ positive integers \cite{hilbert}.

If $N_{dw}  $ is even, i.e., $N_{dw}  = 2K $ with $K   \geq 1 $, there are either $K +1$ upper domains and $K  $ down domains, or $K $ upper domains and $K +1$ down domains. The total number of configurations is then 
\begin{eqnarray}\label{doseven}
  \mathcal{D}(E, M) &=& \binom{N_\uparrow - 1 }{K } \binom{N_\downarrow - 1 }{K-1} + \binom{N_\uparrow - 1 }{K -1 } \binom{N_\downarrow - 1 }{K } \nonumber \\
  &=& \binom{N_\uparrow - 1 }{K-1 } \binom{N_\downarrow - 1 }{K-1} \frac{N_\uparrow -K }{K}  + \binom{N_\uparrow - 1 }{K -1 } \binom{N_\downarrow - 1 }{K -1} \frac{N_\downarrow -K }{K} \nonumber \\
  &=& \binom{N_\uparrow - 1 }{K-1 } \binom{N_\downarrow - 1 }{K-1} \frac{L - 2K }{K } . 
\end{eqnarray}

\begin{figure*}[tb]
\includegraphics[width= 0.8\textwidth ]{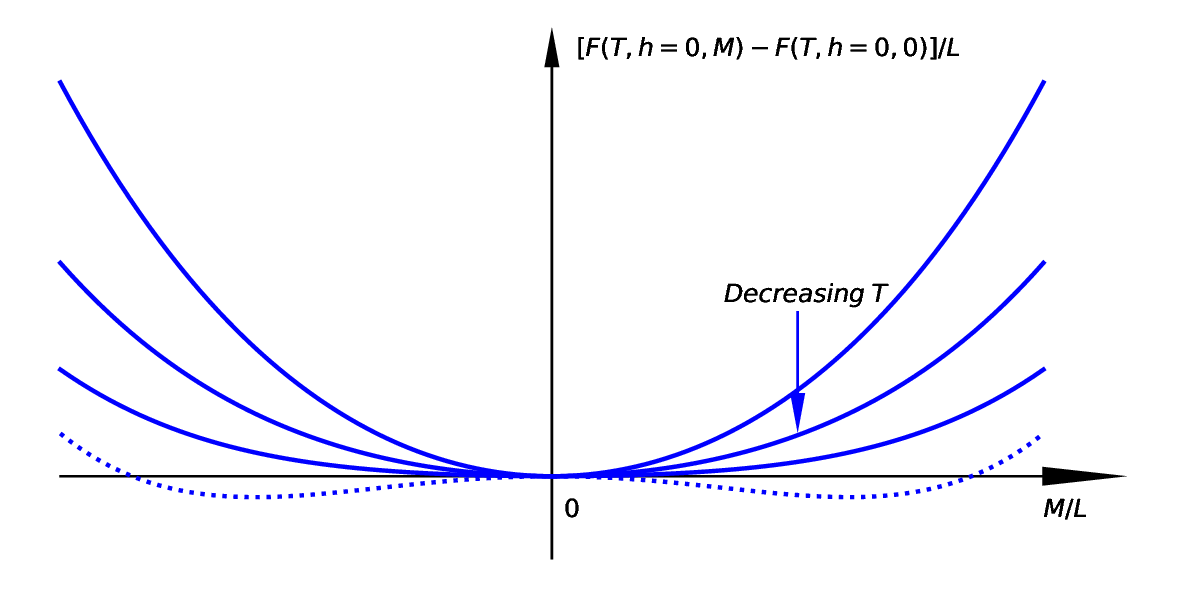}
\caption{(Color online) Sketch of the Landau free energy of the 1D Ising model based on the monotonicity property of the density of states revealed in Fig.~\ref{fig1} and Tab.~\ref{tabledos}. Although the free energy curve becomes increasingly flat as temperature $T$ decreases, its global minimum stays at $M=0$. In other words, the iconic double-well curve (dotted line) of a Landau second-order phase transition—corresponding to symmetry breaking or spontaneous magnetization—can never occur.   }
\label{fig_free}
\end{figure*}

An interesting and useful observation is that the following inequality holds, 
\begin{eqnarray}\label{indos}
  \mathcal{D}(E, M) &\geq & \mathcal{D}(E, M+2), \quad  \quad   0\leq M \leq L - 4 .
\end{eqnarray}
Figure \ref{fig1} and Table \ref{tabledos} demonstrate this point for a lattice of $L = 21$ sites and a lattice of $L=14$ sites, respectively. Plugging this inequality into (\ref{partialz}), we immediately have that in the absence of an external field ($h= 0 $), 
\begin{eqnarray}\label{inz}
  Z(T, h=0, M) &\geq & Z(T, h=0, M+2 ) , \quad  \quad   0\leq M \leq L - 4 . 
\end{eqnarray}
As for the value of $M = L-2 $, we do not have Eq.~(\ref{indos}). But anyway, 
\begin{eqnarray}\label{inz2}
  Z(T, h=0, L-2)& =&  2 e^{\beta(L-3)} + (L-2) e^{\beta(L- 5 )}   \nonumber  \\
  &\geq & e^{\beta (L-1)} = Z( T, h=0, L  ),
\end{eqnarray}
provided $L $ is large enough. Therefore, in the thermodynamic limit, $Z( T, h=0, M\geq 0 )$ is a monotonically decreasing function of $M$. Its maximal value is at $M^* = 0 $ or $1$. This heuristically suggests the absence of spontaneous magnetization of the model. In Fig.~\ref{fig_free}, we sketch the evolution of the curve of the Landau free energy $F(T, h=0, M)=- T \ln Z(T, h=0, M)$ as temperature $T$ decreases. The curve should flatten but can never develop a W shape. 

It is easy to prove (\ref{indos}). If $N_{dw}   =2K -1 $, by (\ref{dosodd}), what we need to prove  is 
\begin{eqnarray}\label{ineq}
  \mathcal{D}(E, M) &=& 2\binom{N_\uparrow -1 }{K -1 } \binom{N_\downarrow -1 }{K-1 } \nonumber \\
    &\geq & 2 \binom{N_\uparrow }{K -1 } \binom{N_\downarrow -2 }{K-1 } = \mathcal{D}(E, M+2).
\end{eqnarray}
Here the point is that as $M \rightarrow M+2$, $N_\uparrow \rightarrow N_\uparrow + 1 $ and $N_\downarrow \rightarrow N_\downarrow - 1 $.
The ratio of the right hand side to the left hand side is 
\begin{eqnarray}
  \frac{N_\uparrow}{N_\uparrow -K +1} \frac{N_\downarrow -K}{N_\downarrow -1} . 
\end{eqnarray}
The difference between the numerator and the denominator is $-(K-1)(N_\uparrow - N_\downarrow +1) $, which is apparently non-positive. Thus the ratio is smaller than or equal to 1. 

If $N_{dw}  = 2K $, by (\ref{doseven}), what we need to prove is 
\begin{eqnarray}
  \binom{N_\uparrow - 1 }{K-1 } \binom{N_\downarrow - 1 }{K-1} &\geq &  \binom{N_\uparrow  }{K-1 } \binom{N_\downarrow - 2 }{K-1} .
\end{eqnarray}
But this  is exactly (\ref{ineq}). 

We have thus proven (\ref{indos}) and in turn (\ref{inz}). 

Here some remarks are in order. First, while the proof of (\ref{indos}) is simple and rigorous, it is somewhat formal and unilluminating. A more insightful proof would entail constructing an explicit injective mapping from the set of configurations with magnetization $M+2$ to those with magnetization $M$. Unfortunately, such a combinatorial construction has thus far eluded us. Second, we point out the trivial global bound:
\begin{align}\label{sumofdos}
  \sum_E \mathcal{D}(E, M) = \binom{L  }{N_\uparrow }  \geq \binom{L  }{N_\uparrow  +1 }  = \sum_E \mathcal{D}(E, M+2),
\end{align}
for $M\geq 0 $. Viewed in this light, inequality (\ref{indos}) is a significantly more refined statement, demonstrating that this monotonicity holds not just globally, but term-by-term at every specific energy level $E$.

So far, we have considered only the open boundary condition. But the same monotonicity property holds for the periodic boundary condition. Under this condition, the interaction energy $E $ in (\ref{eint}) gains an extra term $-s_1 s_L $, and its lowest value is now $-L $. Denote the density of states by $\tilde{\mathcal{D}}(E, M)$ in this case. Now a configuration $\vec{s}$ in the $(E, M)$ class under the open boundary condition, will be in the class $(E+1, M)$ if $E=-(L-1) + 2\times (2K-1)$ for some positive $K$, as in this case its left-most domain and right-most domain are antiparallel; or it will be in the class $(E-1, M)$ if $E = -(L-1) + 2\times  2K $, as in this case its left-most domain and right-most domain are parallel. Therefore, we have
\begin{eqnarray}
  \tilde{\mathcal{D}}(E,M) &=& \mathcal{D}(E+1, M) + \mathcal{D}(E-1, M). 
\end{eqnarray}
Here $E = - L + 4 K $ for $0\leq K \leq \lfloor{L/2} \rfloor $. Because of the monotonic relation (\ref{indos}), we have the monotonic relation  
\begin{eqnarray}
  \tilde{\mathcal{D}}(E, M) &\geq & \tilde{\mathcal{D}}(E, M + 2 ) , \quad \quad  0\leq M \leq L - 4 . 
\end{eqnarray}

\section{Free energy}
In the preceding section, we arrived at the conclusion of no spontaneous magnetization by observing some qualitative characteristics of the density of states. We now transition from this qualitative deduction to a rigorous quantitative analysis. Because we have derived a simple, explicit analytical expression for the density of states, the partition function $Z(T,h)$ in (\ref{z}) is perfectly primed for evaluation. This seemingly formidable summation can, in fact, be carried out exactly, yielding a closed-form expression for $Z(T, h)$. This was the core achievement of Ising's original paper. Ising's result is actually for a finite value of $L $. The thermodynamic limit is obtained trivially thereafter. 
%

Here, we are less ambitious. We bypass the tricky summation problem and aim straight for the thermodynamic limit ($L \rightarrow \infty$). This strategic move allows us to exploit the properties of large systems and apply the maximum-term approximation (or ``saddle-point approximation'')  \cite{maximal} to the partition function, thereby simplifying the calculation significantly. 
Let 
\begin{eqnarray}\label{wmax}
  W_{max} &\equiv & \max_{E, M } W(E, M ). 
\end{eqnarray}
Because the interaction energy $E$ takes $L$ discrete values and the magnetization $M$ takes $L+1$ discrete values, the partition sum in (\ref{z}) consists of at most $L(L+1)$ non-zero terms. Since all statistical weights are strictly positive, the full partition function $Z$ must be bounded below by its single largest term, $W_{max}$, and bounded above by the total number of possible terms multiplied by $W_{max}$. That is, we rigorously have:
\begin{align}\label{boundsofz}
  W_{max} \le Z \le L(L+1)W_{max}. 
\end{align}
Taking logarithm gives 
\begin{eqnarray}\label{uppperandlower}
  \frac{1}{L} \ln W_{max } \leq \frac{1}{L} \ln Z \leq \frac{1}{L} \ln  W_{max} + \frac{1}{L} \ln [L(L+1)] . 
\end{eqnarray}
In the thermodynamic limit ($L\rightarrow \infty$), the ratio $\ln [ L(L+1) ]/L$ vanishes. Therefore, in this limit, the free energy per spin is determined by the maximal term (\ref{wmax}).

\begin{figure*}[tb]
\includegraphics[width= 0.8\textwidth ]{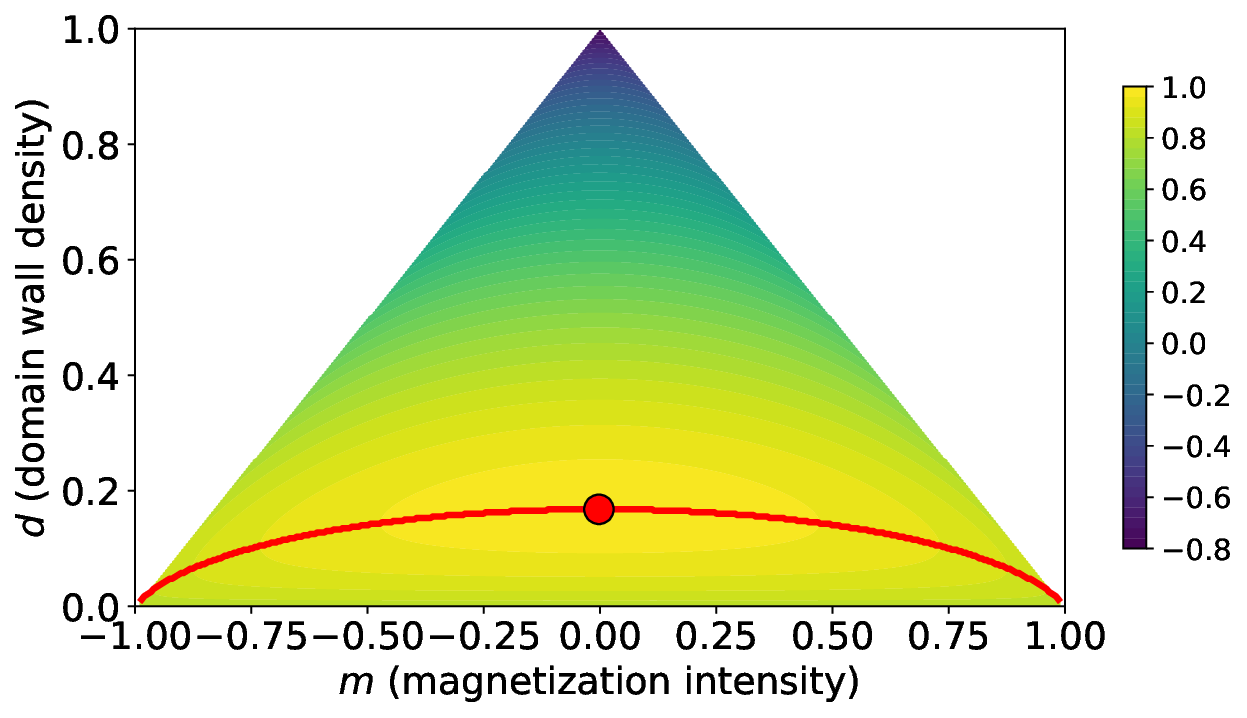}
\caption{(Color online) Heatmap of the function $w(d, m )$ as defined in (\ref{w}). The parameters are $(\beta = 0.8, \alpha = \beta h = 0 )$. The red solid line indicates the maxima of $w $ for fixed $m $, which are points satisfying (\ref{criticalpoint}a). The red dot indicates the global maximum $(d^*, m^*)$, which corresponds to the stable thermodynamic equilibrium state of the model.  }
\label{fig2}
\end{figure*}

We have 
\begin{eqnarray}
  \frac{1}{L} \ln W(E, M) &=& \frac{1}{L}\left [\ln \mathcal{D}(E, M) - \beta E + \alpha M  \right ] . 
\end{eqnarray}
For given temperature $T$ and external field $h$,  we expect that as $L\rightarrow \infty $, this becomes a smooth function of the ratios $E/L$ and $M/L $. Denote this smooth function as $w(d, m)$, where the continuous variables $d $ and $m $ are defined as 
\begin{eqnarray}\label{defofdm}
  \frac{E}{L} &=& -1 + 2d , \quad\quad   \frac{M}{L} = m . 
\end{eqnarray}
Here $d\in [0,1]$ is the density of the domain walls, and $m \in [-1,1]$ is the magnetization intensity. Using Stirling's approximation $\ln N! = N \ln N - N + \mathcal{O} (\ln N )$ for large $N $, we get readily 
\begin{align}\label{w}
w(d, m) &= - d \ln d - \frac{1+m-d}{2} \ln(1+m-d) - \frac{1-m-d}{2} \ln(1-m-d) \nonumber \\
&\quad + \frac{1+m}{2} \ln(1+m) + \frac{1-m}{2} \ln(1-m) - 2\beta d + \alpha m + \beta .
\end{align}
Note that in obtaining (\ref{w}), it does not matter whether we take the expression (\ref{dosodd}) or the expression (\ref{doseven}) of $\mathcal{D}(E, M)  $. The difference between $K$ and $K-1 $ drops out in the thermodynamic limit. In Fig.~\ref{fig2}, we have a plot of the function $w(d, m )$ for a field-free ($\alpha = 0 $) case.

Suppose $w$ attains its maximum at $(d^*, m^*)$. In the thermodynamic limit, $d^*$ and $m^*$ will actually be the mean values of $d$ and $m$, respectively. This is readily seen from the fact that $W(E, M) = e^{L w(d, m)}$. As $L \rightarrow \infty$, this exponential function is exceedingly sharply peaked and can be well approximated by a Gaussian function centered at $(d^*, m^*)$ with a width on the order of $1/\sqrt{L}$. Thus, macroscopic fluctuations vanish, and the most probable state dictates the thermodynamics.

The stationary point $(d^*, m^*)$ is the solution of the equations 
\begin{subequations}\label{criticalpoint}
 \begin{eqnarray}
  \frac{\partial w}{\partial d }   &=& \frac{1}{2} \ln \frac{(1+m - d)(1- m - d ) }{d^2 } - 2 \beta = 0 , \\
  \frac{\partial w }{\partial m } &=&   \frac{1}{2}\ln \frac{(1+m )(1-m - d)}{(1-m)(1+ m - d )} + \alpha = 0 .
 \end{eqnarray}
\end{subequations}
From (\ref{criticalpoint}b), we easily solve $d^* $ in terms of $m^*$, 
\begin{eqnarray}\label{dinm}
  d^* &=&  \frac{1- m^{*2}}{1 + m^* \coth \alpha }  .
\end{eqnarray}
Substituting this back in (\ref{criticalpoint}a), after some straightforward manipulation, we get a simple quadratic equation of $m^*$, and solve 
\begin{eqnarray}\label{mcritical}
  m^* &=& \frac{\sinh \alpha }{\sqrt{\sinh^2 \alpha  + e^{- 4 \beta }}}  .
\end{eqnarray}
Substituting this back in (\ref{dinm}), we get 
\begin{eqnarray}\label{dcritical}
  d^* &=& \frac{e^{-4 \beta }}{\sqrt{\sinh^2 \alpha  + e^{- 4 \beta }} (\sqrt{\sinh^2 \alpha  + e^{- 4 \beta }} + \cosh \alpha  )} . 
\end{eqnarray}
Now we can substitute (\ref{dinm}), (\ref{mcritical}), and (\ref{dcritical}) in (\ref{w}) to obtain the free energy per spin in the thermodynamics limit. It is
\begin{eqnarray}
  f(T, h ) &=& -T \lim_{L\rightarrow \infty } \frac{\ln Z(T, h)}{L} = -T \lim_{L\rightarrow \infty } \frac{\ln W_{max}}{L} = - T w(d^*, m^* ) \nonumber \\
  &=&   -1 - T \ln [\cosh \alpha + \sqrt{\sinh^2 \alpha + e^{-4 \beta }}] . 
\end{eqnarray}
This agrees with the result obtained by the transfer matrix method in \cite{peliti}. 

We see from (\ref{mcritical}) that  in the absence of the external magnetic field ($\alpha \equiv  \beta h = 0 $), $m^*=0$, confirming the absence of spontaneous magnetization. 
Furthermore, as a sanity check of our maximum-term approximation, we evaluate the energy density in this case. By (\ref{dcritical}), we have the domain wall density $d^* = 1/(1+ e^{2\beta})$. Consequently, the energy density is given by [see Eq.~(\ref{defofdm})]
\begin{align}\label{energydensity}
\lim_{L\rightarrow \infty }\frac{\langle E \rangle}{L} &= -1 + 2d^* = -\tanh \beta.
\end{align}
Reassuringly, this expression is in agreement with established textbook results \cite{mattis}, which hold for a finite open chain: $\langle E \rangle = -(L-1) \tanh \beta $. 
This result allows us to easily estimate the thermal fluctuations of the energy in a finite system. The variance of the energy is given by the standard thermodynamic relation $\langle (\Delta E)^2 \rangle \equiv \langle E^2 \rangle - \langle E \rangle^2 =   -\partial \langle E \rangle / \partial \beta$ \cite{peliti,landaulifshitz,mattis, kardar}. Using the exact expression of $\langle E \rangle $, we find $\langle (\Delta E)^2 \rangle = (L-1) \, \text{sech}^2 \beta$. The relative fluctuation is therefore $\Delta E / |\langle E \rangle| = 1/(\sqrt{L-1} \sinh \beta)$. Because this relative fluctuation scales as $1/\sqrt{L}$, it explicitly quantifies the vanishing width of the Gaussian peak of $W(E,M)$ discussed earlier, and provides a clear justification for the maximum-term approximation in the thermodynamic limit.

\section{Landau free energy (in the field-free case, $h = 0 $)}

While (\ref{indos}) and (\ref{inz}) make the conclusion of absence of spontaneous magnetization reasonable, still we want to be more quantitative. Below, we show that the Landau free energy per spin increases monotonically with the magnetization density $m $. 

Now we seek the maximal term for a fixed $M$. Let 
\begin{eqnarray}
  W_{max}^{(M) } & \equiv & \max_E  W(E, M )  .
\end{eqnarray}
Like (\ref{uppperandlower}), we have
\begin{eqnarray}
  \frac{1}{L} \ln  W_{max}^{(M)} &\leq & \frac{1}{L} \ln Z(T, h, M) \leq \frac{1}{L} \ln W_{max}^{(M)} + \frac{1}{L} \ln L  .  
\end{eqnarray}
Therefore, we can again use the maximum-term approximation, and the Landau free energy per spin is 
\begin{eqnarray}
  f(T, h=0 ,  m ) &=& -T \max_d w(d, m )= - Tw(d(m), m ),   
\end{eqnarray}
where $d(m )$ is determined by equation (\ref{criticalpoint}a), i.e., $\frac{\partial w }{\partial d} \big |_{d(m), m}=0 $. In Fig.~\ref{fig2}, the function $d(m)$ is indicated by the red line. We show that $f$ increases with $m \geq 0$ by examining its derivative with respect to $m$. 
We have (essentially an application of the envelope theorem \cite{envelope})
\begin{eqnarray}
  \frac{d f }{d m } &=& -T \left( \frac{\partial w }{\partial m } + \frac{\partial w }{\partial d }\frac{\partial d}{\partial m } \right) \bigg |_{d(m), m } = - T  \frac{\partial w }{\partial m } \bigg |_{d(m), m } .
\end{eqnarray}
After some straightforward calculation, we get a neat expression 
\begin{eqnarray}\label{1stderivative}
  \frac{d f }{d m } &=& -\frac{T}{2} \ln \left [ \frac{(1+m)(1-m - d (m))}{(1-m)(1+m - d (m))} \right ] .
\end{eqnarray}
The difference between the numerator and the denominator in the square brackets is $-2 m d(m) \leq  0$ (Note that $m\geq 0$ by assumption). Therefore, 
\begin{eqnarray}\label{dfdm}
  \frac{d f }{d  m }\bigg |_{m\geq 0 } &\geq &  0 , 
\end{eqnarray}
and the Landau free energy attains its minimum at $m = 0 $, in alignment with (\ref{inz}) and  proving the absence of spontaneous magnetization. 

To further characterize the free energy landscape, we evaluate the second derivative of $f$ with respect to $m$. Differentiating (\ref{1stderivative}) directly gives
\begin{align}\label{2ndderivative}
\frac{d^2 f }{d m^2} & = -T \left[ \frac{1}{1-m^2} -\frac{1-d + m d'}{(1-d)^2 - m^2} \right],
\end{align}
where $d' $ is the derivative of $d$ with respect to $m$. The implicit dependence of $d$ on $m$ is governed by (\ref{criticalpoint}a), which upon implicit differentiation yields the compact relation $d'(m) = -m d/(1 - d- m^2)$. Substituting this result into (\ref{2ndderivative}) allows for a significant algebraic simplification, reducing the second derivative to
\begin{align}\label{2nd2}
\frac{d^2 f }{d m^2} & = \frac{Td}{(1-m^2)(1-d-m^2 )}.
\end{align}
We are particularly interested in its value at $m=0$ as it determines the stability of the zero-magnetization state. At this point, we have $(1-d)/d = e^{2 \beta } $ from (\ref{criticalpoint}a), and thus 
\begin{align}\label{curvature}
  \left. \frac{d^2 f }{d m^2} \right|_{m= 0} = \frac{Td}{1 - d } = T e^{-2/T} > 0.
\end{align}
This strictly positive second derivative confirms that $m=0$ is always a local minimum [actually a global one by (\ref{dfdm})] and never loses its thermodynamic stability, proving again the absence of spontaneous magnetization. Interestingly, this curvature is non-analytic as $T\rightarrow 0 $, which stands in stark contrast to the standard Taylor-expansion assumptions typically made in phenomenological Landau phase transition theory.

\section{Conclusions and discussions}
%

We have revisited the 1D Ising model \cite{1dising, 1dising2} from the Landau perspective, with particular emphasis on the role of density of states in phase transitions \cite{rmp}. Although fundamental quantities such as the partition function admit simple expressions in terms of the density of states, the latter is generally difficult to compute and is therefore rarely a practical starting point. Fortunately, for the 1D Ising model the density of states can be obtained exactly by following Ising’s original combinatorial approach. Even more strikingly, this density of states exhibits a monotonicity property that provides a strong heuristic argument against the possibility of spontaneous magnetization at any finite temperature. While this is a well-known result, we believe the density of states perspective offers fresh insight into the problem. It is also interesting that this century-old model still allows for new understanding.


Contemporary textbooks typically solve the 1D Ising model using the transfer-matrix method \cite{peliti, mattis, kardar}, which is indeed the simplest, most elegant, and most revealing technique.  However, they often fail to mention that the transfer matrix method is not the original method used by Ising himself to solve his model. Actually, the transfer matrix method was invented and applied to the Ising model by Kramers and Wannier in 1941 \cite{kramers}---sixteen years later than Ising's 1925 paper. 

Ising's original approach is combinatorial in style. His key observation is that, a 1D lattice consists of alternating domains and energy penalty comes only from domain walls. This insight makes the counting and summation of the configurations in the partition function possible. His solution, although not so elegant and powerful as the transfer matrix method, is still ingenious and highly instructive. It is rather unfortunate that this solution does not appear in any of the textbooks we surveyed. 



We borrowed from Ising the technique to calculate the density of states. Our approach departs from his at the next step of calculating the partition function. While he skillfully summed up all the terms for a finite lattice, we targeted directly an infinite lattice by using the maximum-term approximation. Our approach is admittedly less elegant than either Ising’s finite-$N$ exact solution or the transfer-matrix method, as it applies strictly only to infinitely large systems. Nevertheless, it demonstrates the power of the maximum-term approximation, which lies at the heart of statistical mechanics. 


Finally, a few words about the pedagogical value of the paper are in order. Technically, the mathematics is elementary and all the calculations are accessible to a competent undergraduate student. The material can thus serve as an exercise for undergraduate and graduate students. In our opinion, it is appropriate to solve the 1D Ising model in the class by using the transfer matrix method and then ask the students to redo it using the Ising and present methods. Through this exercise, students can review many fundamental concepts, such as density of states, partition function, maximum-term approximation, and Landau free energy. Physically, the problem provides a good case of studying the presence or absence of spontaneous symmetry breaking from the  Landau free energy perspective. Textbooks usually introduce the Landau free energy in the framework of Landau phase transition theory \cite{kardar} and assume that at low temperatures, the free energy curve changes from a U shape to a W shape. Now we see that the 1D Ising model is a counter-example, as the free energy curve remains U-shaped regardless of how low the temperature drops. We also note that while the Landau free energy and the 1D Ising model are both standard textbook topics, the application of the former to the latter is rarely studied beyond the mean-field approximation \cite{tanaka}, which erroneously predicts spontaneous magnetization below some critical temperature. Now the gap is filled.  

\section*{Acknowledgments}

The authors are grateful to B. R. Que for posing the question which motivated the study and Q. H. Liu, K. Jin and Y. Xiang for their helpful comments.


\section*{References}

\begin{thebibliography}{99}

\bibitem{Ising1925}
Ising E 1925 Beitrag zur Theorie des Ferromagnetismus \textit{Z. Phys.} \textbf{31} 253

\bibitem{english}
An English translation of Ising's original paper can be found here: \url{https://www.hs-augsburg.de/~harsch/anglica/Chronology/20thC/Ising/isi_fm00.html}

\bibitem{peliti}
Peliti L 2011 Statistical Mechanics in a Nutshell (Princeton: Princeton University Press)

\bibitem{landaulifshitz}
Landau L D and Lifshitz E M 1986 Statistical Physics Part 1 (Singapore: World Scientific) 

%

\bibitem{landau}
Landau L D 1936 The theory of phase transitions Nature \textbf{138} 840

\bibitem{russian}
Semenov S and Rubtsov A N 2024 Landau free energy of small clusters beyond mean-field approach  Phys. Rev. E \textbf{110} L012101 

\bibitem{hilbert}
This standard combinatorial problem now has a standard solution known to all: the ``stars and bars'' method. It is interesting to note, however, that this technique appears to have been unknown to David Hilbert. He considered the problem in his book \emph{Theory of Algebraic Invariants} but solved it using mathematical induction, which is more tedious.

\bibitem{maximal}
Hill T L 1987 An Introduction to Statistical Thermodynamics (New York: Dover Publications)

\bibitem{envelope}
Dominguez T and Mourrat J C 2023 Statistical mechanics of mean-field disordered systems: A Hamilton-Jacobi approach arXiv:2311.08976

\bibitem{1dising}
 Seth S  2017 Combinatorial approach to exactly solve the 1D Ising model Eur. J. Phys. \textbf{38} 015104

\bibitem{1dising2}
Magare S, Roy A K  and Srivastava V 2022  1D Ising model using the Kronecker sum and Kronecker product Eur. J. Phys. \textbf{43} 035102

\bibitem{rmp}
Kastner M 2008  Phase transitions and configuration space topology Rev. Mod. Phys. \textbf{80} 167

\bibitem{mattis}
Mattis D C and Swendsen R H 2008 Statistical Mechanics Made Simple (Singapore: World Scientific)

\bibitem{kardar}
Kardar M 2007 Statistical Physics of Fields (Cambridge: Cambridge University Press)

\bibitem{kramers}
Kramers H A and Wannier G H 1941 Statistics of the Two-Dimensional Ferromagnet. Part I Phys. Rev. \textbf{60} 252 

\bibitem{tanaka}
Tanaka T 2002  Methods of Statistical Physics (Cambridge: Cambridge University Press)



\end{thebibliography}
\end{document}